\documentclass[twocolumn,showpacs,preprintnumbers,amsmath,amssymb,nofootinbib]{revtex4}
\usepackage{graphicx}
\usepackage{dcolumn}
\usepackage{bm}
\begin{document}

\title{ 373 K Superconductors }

\author{Ivan Zahariev Kostadinov}

\affiliation{\textit{373K-SUPERCONDUCTORS},
{\textit{Private Research Institute}}}
\date{\today}

\begin{abstract}  Experimental evidence of superconductors with critical temperatures above $373\:K$ is presented. In a family of different compounds we demonstrate the superconductor state, the transition to normal state above $387\:K$, an intermediate $242\:K$ superconductor, susceptibility up to $350\:K$, $I-V$ curves at $4.2\:K$ in magnetic field of $12\:T$ and current up to $60\:A$, $300\:K$  Josephson Junctions  and Shapiro steps with radiation of $5\:GHz$ to $21\:THz$, $300\:K$ tapes tests with high currents up to $3000\:A$ and many $THz$ images of coins and washers. Due to a pending patent$\footnote{For inquiries about the USPTO application contact Thomas W. Humphrey, Partner, Wood Herron and Evans LLP.}$, the exact chemical characterization and technological processes for these materials are temporarily withheld and will be presented elsewhere.
\end{abstract}
\maketitle
\section{Introductory remarks}

The search for superconductors with increasingly higher  $T_{c}$  has started with the discovery of Kamerlingh Onnes\cite{1}, who on April 08, 1911 observed the electrical resistance of pure mercury vanish at liquid $He$ temperature of $4.2\:K$. After the 1986 discovery of superconductivity in perovskite-type oxides by Bednortz and Mueller\cite{2}, a new age in the quest for room temperature superconductivity has begun. In a short span of time till 1993, the critical  temperature increased up to $163\:K$, in materials under external pressure. As these events are well known to the community, we will refer the reader to  the excellent reviews and analysis of the observed phenomena\cite{3}. Other very recent advancements worth mentioning are in systems at extremely high pressures\cite{HD3.4Mbar,203K}.

As many others, our group started working on high-temperature superconductivity in 1987. Some of the topics we have been interested in are listed in the references\cite{I-V,Bi,TL,Fano,Vanadium,Lifshitz,Nd,Hg}. The present author was active on the broad subject\cite{85}, and continued working on the synthesis of various superconductors over the years.

In this communication we focus our attention on novel superconductor materials we have synthesized and characterized with transition temperatures in the range of $\textbf{373\:K = 100\:C}$ degrees and even well above, because of their obvious importance.
Below we present experimental results for selected samples of these materials obtained in the previous century, marked as type \textbf{A, B, C, D=E} and some recent ones.

\begin{itemize}
  \item Resistance temperature dependence of the sample \textbf{A} in the range between $350$ and $388\:K$ with a zero resistance state and a transition point at \textbf{387\:K}.
  \item  AC magnetic susceptibility of a sample type \textbf{B} with intermediate critical temperature \textbf{${T_{c} = 242\:K}$}.
  \item Magnetic susceptibility data from the twin sample \textbf{E}, collected during several runs on different temperatures in a Quantum Design system. The curves  show negative susceptibility up to $5\:T$ in the range from $60\:K$ to $350\:K$. The system does not reach normal state due to a restriction in temperature ranges of the experimental setup.
  \item $I-V$ data at \textbf{4.2\:K} of the sample \textbf{D} that demonstrates a typical superconductors behavior in high magnetic field of up to  \textbf{12\:T} and current of up to \textbf{60\:A}. It proves existence of superconductivity in the twin sample to the one studied in the Quantum Design for magnetic susceptibility (sample \textbf{E}). The interpretation of the observed curves indicates possible presence of magnetic-flux avalanches.
  \item The gap in the energy spectrum was evaluated via measuring the tunneling current at \textbf{300\:K} of a sample of the type \textbf{C}. The gap was found to be about \textbf{15\:mV} in several measurements, the one shown is based on 9000 data points its lowest current being  6.49\:pA. The waiting time is the reason to collect larger current values at the same voltage due to the existence of parallel super-current paths in the pellet.
  \item A Josephson Junction (JJ)\cite{JJ} is shown, with several other setups marking the broad range of frequencies from about \textbf{5\:GHz} up to \textbf{21\:THz} involved in the Shapiro steps\cite{SSt} and the JJ mesa-type \textbf{THz}  waves generation.
  \item \textbf{THz} waves images of coins and small washers are also displayed with tiny features indicating few \textbf{THz}  waves presence.
  \item Experiments were performed by passing $125\:A$ current via: a) a superconductor tape (black) and simultaneously in series with b) a copper tape of the same size and thickness at room temperature of \textbf{$300\:K$ }. The copper tape started burning after about $5$ seconds. The technical current density was about \textbf{$10,000\:A/cm^{2}$}, which is comparable with industrial high current transport superconductors.

\end{itemize}

 The data presented here are based on the above-mentioned samples of composite materials, of types \textbf{A, B, C, D=E} superconductors and  some recently produced ones. These marked samples (with the exception of sample \textbf{B}) are all with $T_c$ in the broad vicinity of $373\:K$. Other compounds with slightly higher $T_c$ were synthesized and tested but data is not presented here as they were not fully characterized. As a perspective, our decades-long studies include compositions of elements from several major groups in the periodic table, many of them with lower or higher critical temperatures, and many of them non-superconductors at all. In this exhaustive investigation we were initially guided by our theoretical ideas and practical experience in the HTS materials. The details of these novel insights we have gained over our study, which have led to the current materials with room-temperature $T_c$'s, will be shared in upcoming publications.

\section{The DC resistance transition}
\label{Sec1}
\begin{figure}[]
\includegraphics[width=\columnwidth]{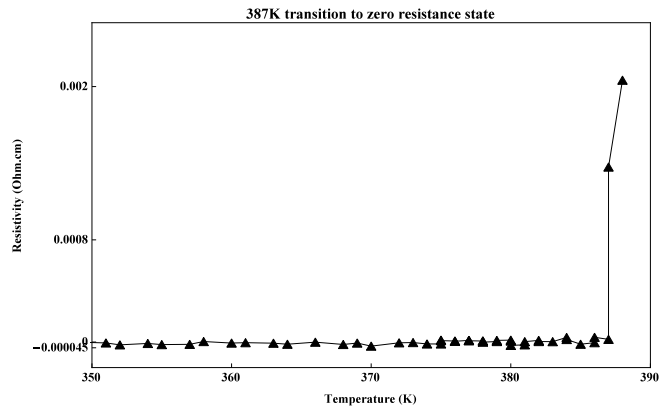}
\caption{ The DC resistance temperature dependence of an early $\textbf{A}$ type sample. The measuring current was $0.5\:A$ and the resistivity is in $Ohm.cm$. The $\textbf{Tc = 387\:K or 114\:C}$ and the lowest measured voltage is in the few micro Volts range.}
\label{fig.1}
\end{figure}

First we describe the DC resistance temperature dependence of a macroscopical sample across it and demonstrate the critical temperature $T_c$ of the transition from superconductive to normal state. In $Fig. {\ref{fig.1}}$ we show that the sample is still in zero resistance state above the boiling water temperature of \textbf{$100\:C$} and that the transition occurs at about $387\:K$ or $114\:C$. The recorded data obtained from a sample of the composition \textbf{A} type group are displayed. Many of the initially studied samples were compacted to form circular pellets at a pressure of about $ {5-10\:kbar}/{cm^{2}}$  of about $1\:cm$ diameter and about $2\:mm - 3\:mm$ thickness with four contacts attached made of In-Ga eutectic composition. Other size pellets were also pressed at a pressure of ${5-10\:kbar}/{cm^{2}}$ and studied. The temperature was measured with a platinum $100\:Ohms$ thermometer attached to the sample inside the sample holder. There were two other thermometers and a heater built in our standard closed cycle cryostat with lowest temperature of 10K. Measurements in the range from $350\:K$ to $390\:K$ are shown demonstrating a well-visible zero resistance state and a sharp transition at $387\:K$.

\section{ $T_{c}=242\:K$ recorded in the magnetic susceptibility of sample B}
\begin{figure}[b]
\includegraphics[width=\columnwidth]{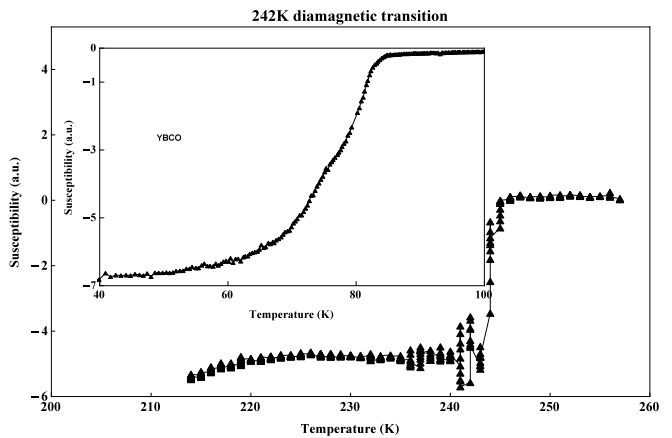}
\caption{The susceptibility of a  \textbf{B} type sample  in relative units showing a critical temperature of \textbf{242\:K}. In the inset is the $Y_{1}Ba_{2}Cu_{3}O_{7-x}$ calibration curve of the coil.}
\label{fig.2}
\end{figure}

In this section we show data from a sample with a different composition \textbf{ B} type, as a milestone mark,  studied among many others on the way toward the above \textbf{100 C}  superconductors presented in the first section.
The magnetization temperature dependence is the second fundamental characteristic of the superconductor state, which together with the resistance makes the two observations proving the presence of superconductivity in a macroscopical sample. The magnetization of a sample of \textbf{B} type composition was studied in the same closed cycle cryostat described in Section \ref{Sec1}, with a standard three-coil system. The coils were calibrated with a sample of $Y_{1}Ba_{2}Cu_{3}O_{7-x}$ producing strong signal of the Lock-in amplifier operating at a frequency of $137\:Hz$. The $YBCO$ sample was otherwise not specifically characterized. From the graph in $Fig. {\ref{fig.2}}$ we clearly observe that the amplitude of the transition signal variation of the sample having a transition temperature of $242\:K$ is of the same magnitude as the $Y_{1}Ba_{2}Cu_{3}O_{7-x}$ system used for calibration.

All measured samples in the susceptibility setup were cut in an approximately rectangular 2x2x4 mm shape. A typical discrepancy of about $4-5\:K$ was observed of the critical temperatures of the cooling and warming regimes due to the position of the coil and the thermocouple above the cooler. This difference is absent if the thermal stabilization is applied at every few degrees.
The number $242\:K$ was selected as the last point before the sharp increase due to the transition from a superconductor state to a metallic type of resistance.  This set of data displayed in the graph below is included here to show that the Room Temperature Superconductors (\textbf{RTS}) were found in a process including synthesis of a large amount of various superconductor compositions with intermediate transition temperatures.

\section{Negative magnetization at $B = 5\:T$ and $ T = 350\:K$}
\begin{figure}[]
\includegraphics[width=\columnwidth]{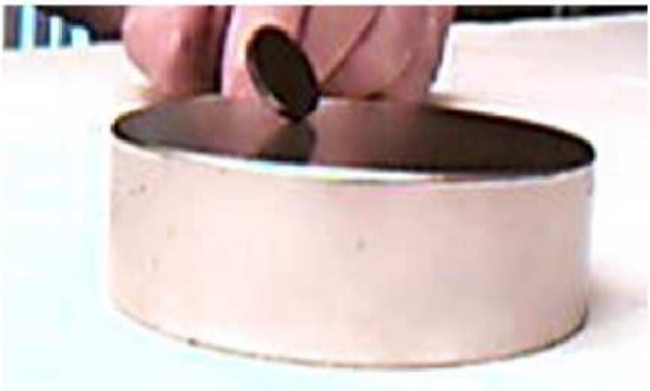}
\caption{ \textbf{RTS} disk of a diameter about $2\:cm$ and about $3.4\:mm$ tick is on the surface of a 3 inch diameter permanent magnet with a  $B \simeq 1.4\:T$ field in the center.}
\label{fig.3}
\end{figure}

Before presenting the data demonstrating the magnetic susceptibility, we are posting a photo of a \textbf{RTS} sample positioned on a $8\:cm$ diameter $Neodimium$ based permanent magnet with magnetic field in the center of about $1.4\:T.$ As with any diamagnet (i.e. copper), the superconductor sample is attracted to the magnet, but being placed in the center of its surface, strong surface super-currents are created shielding the interior of the sample from the magnetic field. If the field is between the first and second critical magnetic field values there may be partial penetration in the form of fluxoids - single flux quanta (or a bunch of them), which could be forming a regular or irregular Abrikosov lattice{\cite{AA}}. In our case there are two stable positions of the disk of about $2\:cm$ diameter: one is horizontal and the other is vertical. In the horizontal position the sample, barely laying on the surface, slides away from the strongest field to the edge as a place with highest gradient of the field. In the vertical position it also tends to escape the center and floats to the edge at a tiny distance above (or below if suspended) the magnet surface. As it is presented in $Fig. {\ref{fig.3}}$, around halfway to the edge, the \textbf{RTS} disk stays slightly tilted and it may stay tilted, or oscillate like a metronome.

\begin{figure}[]
\includegraphics[width=\columnwidth]{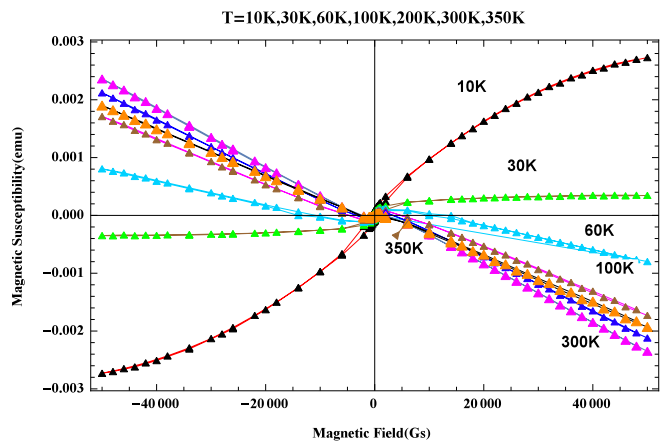}
\caption{The magnetization (-5\:T,+5\:T) scans at selected several temperatures from $10\:K$ to $350\:K$ show a transition from paramagnetic to diamagnetic behavior at room temperatures and higher. An upturn begins to show at $350\:K$. The same sample was in superconductor/diamagnetic state at $4.2\:K$ and $12\:T$ magnetic field from $0$ up to $6\:A$ current(see sec.5)}
\label{fig.4}
\end{figure}

The magnetic field effect on some substances is to induce a small attractive force, which is measured with the diamagnetic susceptibility, studied here in order to find the magnitude of the induced diamagnetic polarization. The data were collected from a sample \textbf{ E} type. The data show only the characteristic  for superconductors negative susceptibility in the range between $60\:K$ and  $350\:K$, as  the transition to a normal metallic state has not been reached since it takes place at higher temperatures, close to or above $400\:K$. The data demonstrate sensitivity in the vicinity of the transition to the normal state. There is a smaller value of the susceptibility at $5\:T$ at $350\:K$, when nearing the transition point (in the range of $400\:K$) as demonstrated in $Fig. {\ref{fig.4}}$ by the less negative magnetization at $350\:K$, when compared to the $300\:K$ curve.

\begin{figure}[]

\includegraphics[width=\columnwidth]{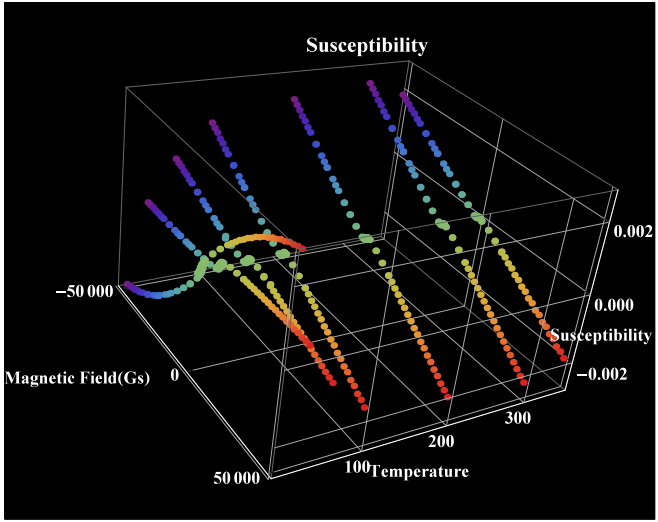}
\caption{The magnetization (-5\:T,+5\:T) scans at the selected temperatures from $10\:K$ to $350\:K$ show a transition from paramagnetic to diamagnetic behavior starting at $60\:K$ and seen at room temperatures and higher. An upturn begins at $350\:K$.}
\label{fig.5}
\end{figure}

The samples \textbf{ D} and \textbf{E } type were two different pieces of a larger one with a specific synthesis history and with critical transition temperature close to $400\:K$ - a temperature not available in the measuring system from Quantum Design used in the experiment.
Next we show a $3D$ graph  $Fig. {\ref{fig.5}}$ of the actual recorded data points.

\begin{figure}[]

\includegraphics[width=\columnwidth]{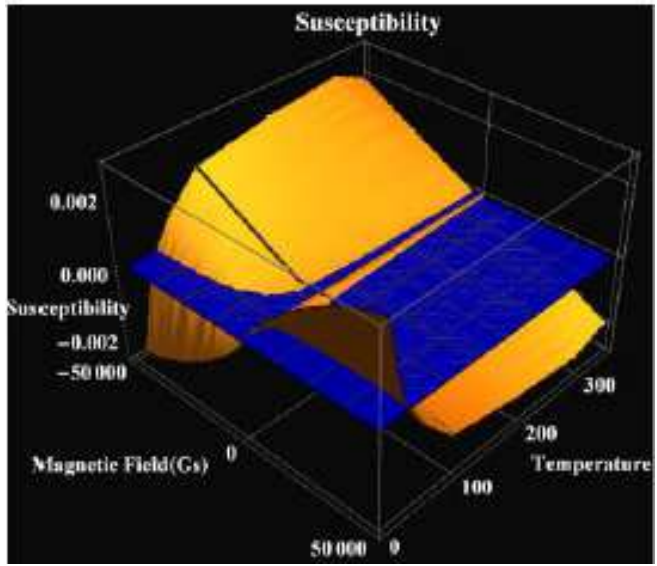}
\caption{The magnetization (-5\:T,+5\:T) scans at selected temperatures from $10\:K$ to $350\:K$ show in blue a transition from paramagnetic to diamagnetic behavior from $60\:K$ to room temperatures and higher. An upturn of the yellow surface begins to show at $350\:K$ the normal state being in the 400K range}
\label{fig.6}
\end{figure}

 The transition back to normal state is in the range of $400\:K$ and could not be reached by the system. We show also a 3D  graph, where the data points are connected with yellow surface. The zero magnetization is marked with a blue plane. The paramagnetic to diamagnetic transition is seen $Fig. {\ref{fig.6}}$  at the crossing lines of the two surfaces-yellow and blue one.

\begin{figure}[]

\includegraphics[width=\columnwidth]{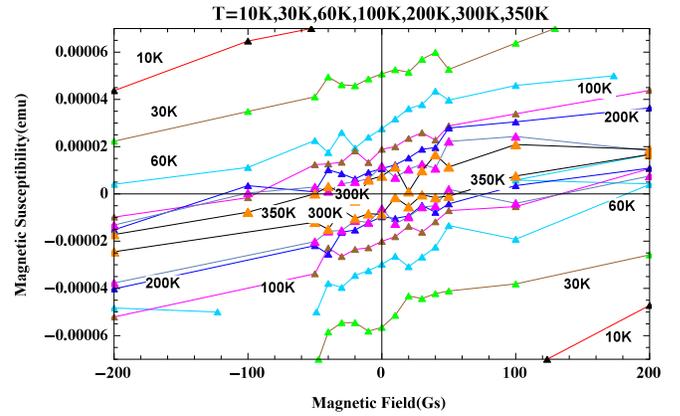}
\caption{The magnetization (-5T,+5T) scans at selected temperatures $10\:K$ to $350\:K$ show a transition from diamagnetic to paramagnetic in the low magnetic  fields region near the origin from $10\:K$ to room temperatures and higher.}
\label{fig.7}

\end{figure}

For a superconductor sample with higher $T_{c}$ the second critical magnetic field is expected to be much higher then $5\:T $ and for this reason a direct downturn transition at lower temperatures ($10\:K$ and $30\:K$) was  not observed (it was found that 12T at 4.2K turn it in diamagnetic state-see sec.5). We first present all the data in a single graph to show the specific for this sample features at low magnetic field.
For all curves, one of the loops near $B = 0$ magnetic filed shows a small initial part of diamagnetic susceptibility expanded here$Fig.{\ref{fig.7}}$  to demonstrate it. All graphs  present a peculiar behavior showing a transition from positive magnetization to a negative one. Except for the $10\:K $ and $30\:K$ all pass another zero becoming negative as seen in the first graph before reaching $5\:T$. To make clear how the loop is recorded we mention that for the $60\:K$ run it starts at $B = 5\:T$ negative, is positive near zero, turns negative at smaller negative values of $B$, then is positive for negative $B = -5\:T$ and returns to zero being negative i.e. near zero the sample is diamagnetic $Fig.{\ref{fig.7}}$ coming from negative values and paramagnetic from positive values of $B$.
 It can be stated with certainty that at higher then $5\:T$ field there will be a downturn at $10\:K$ and $30\:K$ due to the existing superconductivity of the same sample at higher magnetic fields(see next sections).

Further, at $60\:K$ and  all higher temperatures up to $350\:K$, all curves have negative magnetization showing a low-temperatures diamagnetic behaviour, then a paramagnetic magnetization is seen and again downturn diamagnetic at higher magnetic fields, which stays  diamagnetic up to $5\:T$. Moreover, one can trace various lines of transitions in the zero magnetization $B-T$ plane, if the data points are connected to form a 3D surface (see the 3D graph).
 The $\emph{specific synthesis history}$ of the twin samples \textbf{D} and \textbf{E} could be related to the transition from diamagnetic near $B=0$ to paramagnetic and later again diamagnetic as seen on the 3D graph.
It remains to demonstrate that there are samples with transition temperatures higher then the one shown of $387\:K$ and we will give further such records of the magnetization. When the samples of the \textbf{D} and \textbf{E} types are overheated above the \textbf{${T_c}$} in a room temperatures environment to about  $150\:C$ (well above $T_{c}$ ) the magnetization undergoes successive transitions from a superconductor state at room temperature to a non-superconductor one if over-heated well above $T_{c}$ and  well above the boiling point of the water -  like  $150\:C$.

\section{I-V at 12T and 4.2\:K up to 60\:A}

\begin{figure}[b]
\includegraphics[width=\columnwidth]{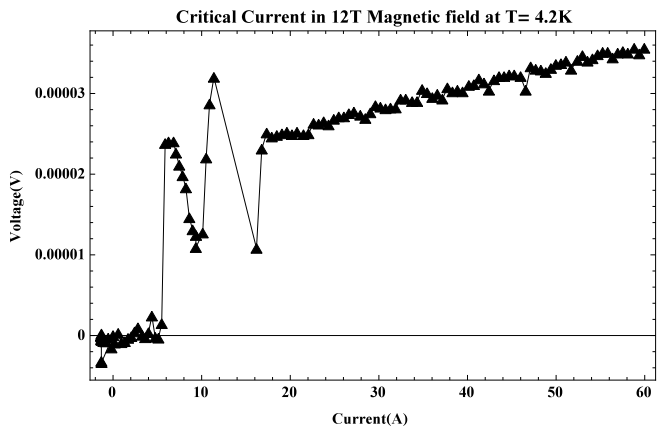}
\caption{I-V curve of the sample  \textbf{ D} at \textbf{4.2\:K} in \textbf{12\:T} magnetic field. The critical current density is about $38\:A/cm^{2}$.}
\label{fig.8}
\end{figure}

The critical current of sample \textbf{E} (same as \textbf{D}) was studied in liquid He $4.2\:K$ and a high magnetic field of $12\:T$. These results are listed next. The previous graph was of sample \textbf{E} showing the magnetization being negative in up to $5\:T$ magnetic field and up to $350\:K$ temperature. Here we study its twin sample \textbf{D} in high magnetic field of $12\:T$ under strong DC current sweep in a heavy experimental setup used for industrial samples serial testing. The sample was in liquid helium at $4.2\:K$ at all times (because of the experimental setup to reach and maintain the high magnetic field) with attached 4 high current contacts of $Sn-Ag$ composition and the field of  \textbf{12\:T} was applied continuously.
From $Fig. {\ref{fig.8}}$  it can be observed that for currents up to about $6\:A$ the sample was in a superconductor state. Comparing this observation with the data of the Quantum Design magnetic susceptibility we come to the conclusion that the \textbf{12\:T} high field was enough strong to overturn the positive susceptibility to negative one in liquid $He$. This observation is \textit{proving our conjecture} from the previous section that the transition at the origin of the \textbf{3D} graph was starting from a superconductor  and became a paramagnetic metal driven by the increase of the magnetic field and with further increase will become again diamagnetic. In the $I-V$ curve we observe a sharp current variation with a remarkable decrease of the current $Fig. {\ref{fig.8}}$. It is probably due to magnetic flux  avalanches, when the magnetic field penetrates the sample. Next we will show that samples with high critical temperature similar to  the \textbf{D, E} types are undergoing transitions from superconductor to normal state in the range of $400\:K$ temperatures.

 \section{Room temperature gap in the energy spectrum from the tunneling $I-V$ curve}
\begin{figure}[]
\includegraphics[width=\columnwidth]{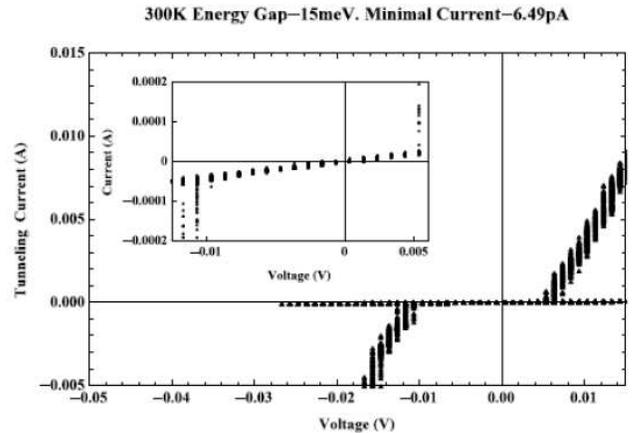}
\caption{The $9000$ data points show presence of a tunneling gap in the sample at \textbf{300\:K} of about \textbf{15\:mV} as seen in the inset. The vertical current jumps at fixed voltage values are similar to the zero voltage current jump in the Josephson junctions and show presence of zero resistance in the sample at any fixed voltage $V$ as  seen from $\frac{\Delta V}{\Delta I} = 0$}.
\label{fig.9}
\end{figure}

The energy spectrum gap (the energy necessary for breaking the pairs) is an important characteristic of any superconductor. Here we report the experimental data from sample \textbf{C} type measured at \textbf{300\:K}. The tunneling $I-V$ curve of a sample type \textbf{C} with contacts was measured with a Keithley pico-Amper meter and the data collected were at a very  close range between each other. The sample was a pellet of about \textbf{1.26 cm} diameter and about \textbf{3 mm} thick with \textbf{4} gold contacts evaporated parallel to each other. The lowest current measured was $6.49\:pA$. Because of the relatively long collection time at any fixed voltage several larger values of the current were collected  along parallel super-current pathways at this particular voltage. It clearly demonstrates superconductivity of the sample at $300\:K$ (Fig. {\ref{fig.9}}) since the local resistance was evidently zero $(\frac{\Delta V}{\Delta I} = 0)$.

A remark about the near zero values of the current at voltages outside the gap is due and it is evidently related to the super-current parallel pathways mentioned about the vertical current lines. More can be said and some theories accounting for it may be mentioned later if there is interest in these phenomena (not restricted to the presented graphs).

\section{Josephson Junctions $I-V$'s}
\begin{figure}[]
\includegraphics[width=\columnwidth]{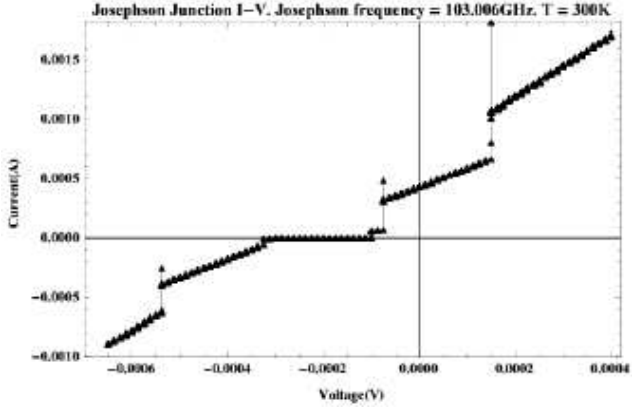}
\caption{Josephson Junction $I-V$. The Josephson frequency is $\textbf{103.006\:GHz}$ at $\textbf{T = 300\:K}$ as seen from the Shapiro steps voltage width.}
\label{fig.10}
\end{figure}

 In this section we will present several $I-V$'s of various samples and other characteristics verifying the superconductivity at $\textbf{300\:K}$ and higher temperatures. We show the room temperature $I-V$ $Fig. {\ref{fig.10}}$  of a mesa-type Josephson Junction (JJ)\cite{JJ} of a size of $0.28\:cm$ acting as a resonator selecting the generated frequency we observed.

\begin{figure}[]
\includegraphics[width=\columnwidth]{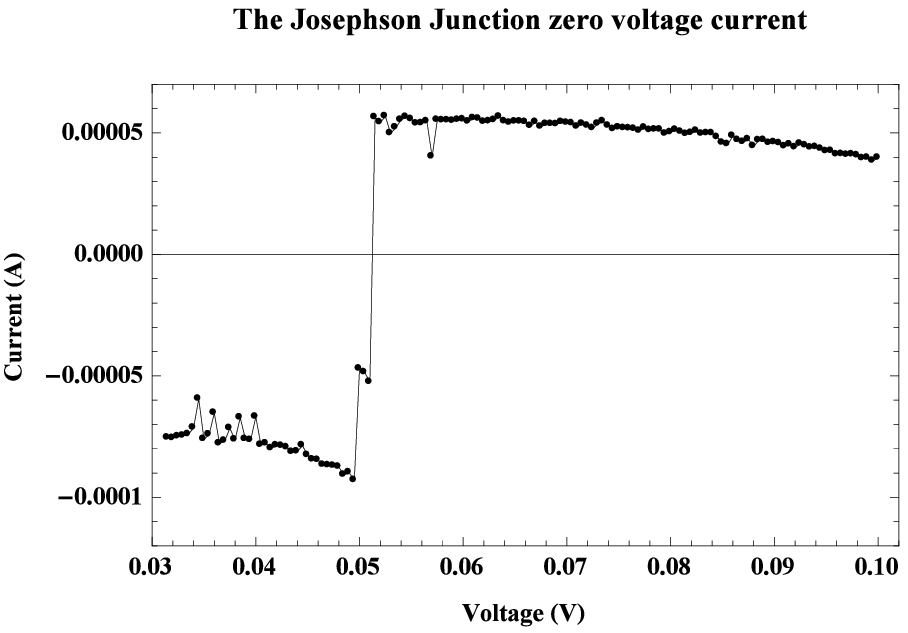}
\caption{Josephson Junction $I-V$ with  zero voltage  super-current magnitude of about $0.1494\:mA $ shifted by a contact voltage $0.0495\:V$ at $\textbf{T = 300\:K}$.}
\label{fig.11}
\end{figure}

In $Fig. {\ref{fig.11}}$ we list the zero voltage Josephson supercurrent jump of about $0.1494\:mA $. Due to the contacts it is shifted to $0.0495\:V$, which is characteristic of the gold contact and the material.

\begin{figure}[]
\includegraphics[width=\columnwidth]{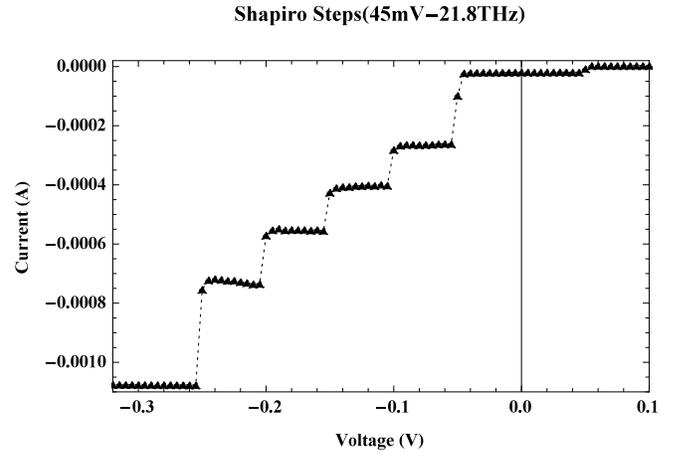}
\caption{Josephson Junction $I-V$ with first current jump up magnitude of about $0.355\:mA $ shifted, the first Shapiro step being about  $45\:mV$  equivalent to a frequency of about $21.762\:THz$ at room temperature $\textbf{T = 300\:K}$.}
\label{fig.12}
\end{figure}

In a different experiment we applied a $10\:GHz$ source and found well-formed Shapiro steps\cite{SSt}. We have often observed Shapiro steps of high frequency created by the mesa-type Josephson Junctions (JJ) with a resonant frequency related to the resonator formed by the circuit. In $Fig.{\ref{fig.12}}$  we show such a JJ $I-V$ with a frequency of $21.762\:THz$.

\begin{figure}[]
\includegraphics[width=\columnwidth]{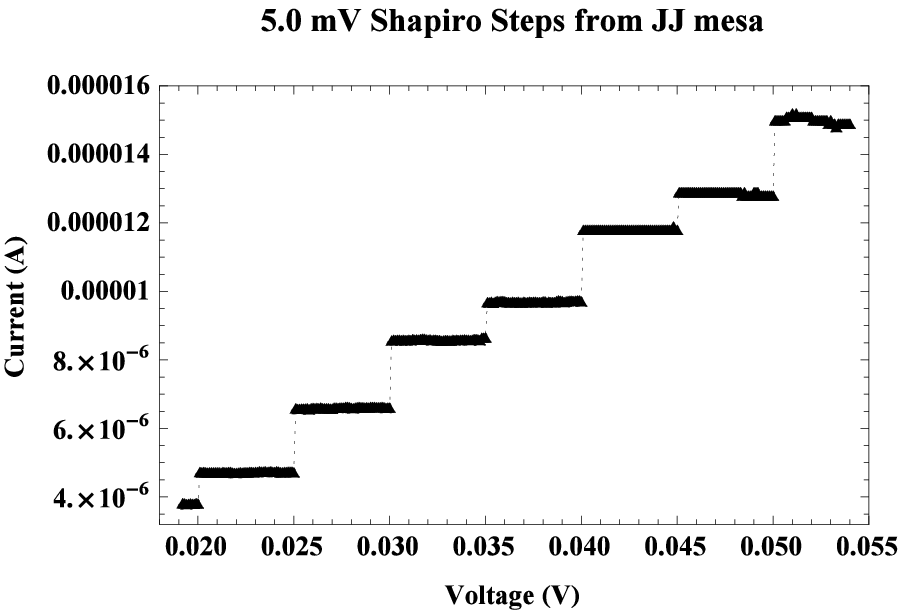}
\caption{Josephson Junction $I-V$ with the first Shapiro step being about  $5.0\:mV$  equivalent to a frequency of about $2.418\:THz$ at room temperature $\textbf{T = 300\:K}$.}
\label{fig.13}
\end{figure}

Other type of Shapiro steps were observed in a broad spectrum of frequencies with the mesa-type of JJ like the ones shown above in $Fig.{\ref{fig.13}}$.

\begin{figure}[]
\includegraphics[width=\columnwidth]{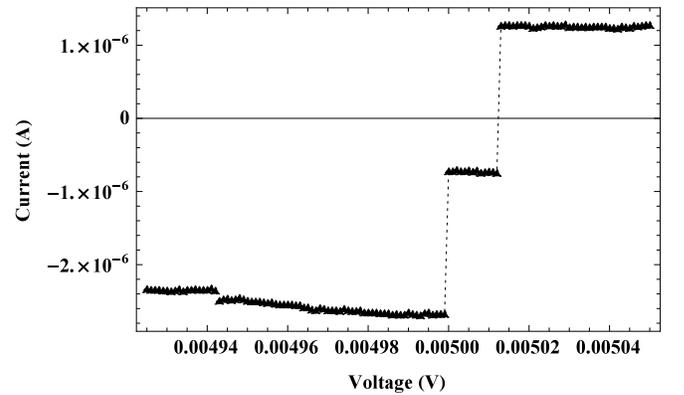}
\caption{Josephson Junction $I-V$ with a single Shapiro step of about  $12\:microV$  equivalent to a frequency of about $5.803\:GHz$ at room temperature $\textbf{T = 300\:K}$.}
\label{fig.14}
\end{figure}

Yet another, different single Shapiro step is observed with the similar mesa-type of JJ like the ones in the previous figure of a different frequency $Fig. {\ref{fig.14}}$.

\begin{figure}[]
\includegraphics[width=\columnwidth]{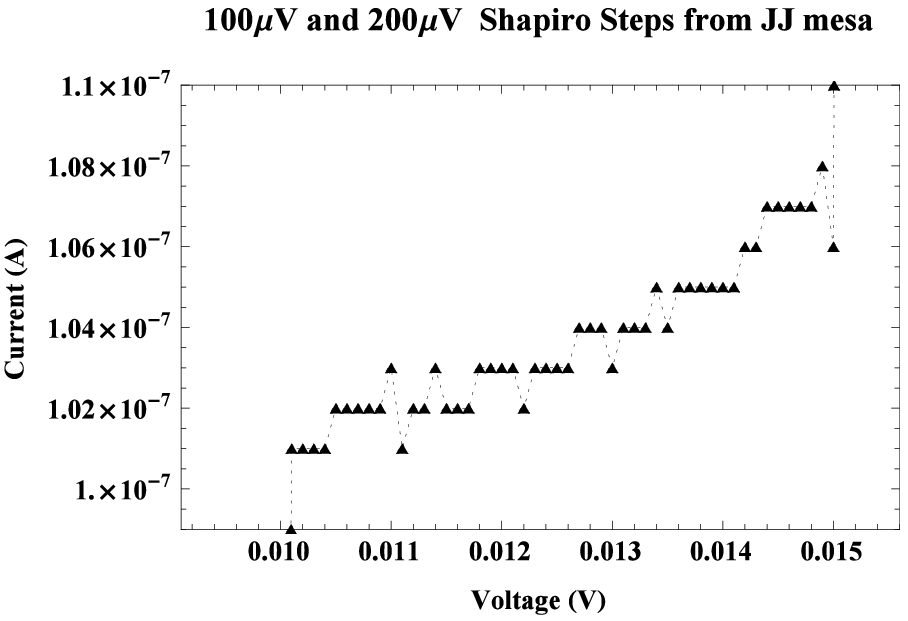}
\caption{Josephson Junction $I-V$ with different Shapiro steps within one long step of about $4.7\:mV$ , the smallest being of $100$ and $200\:microVolts$  equivalent to a frequency of about $48.36\:GHz$ and $96.72\:GHz$ at room temperature \textbf{T = 300\:K}.}
\label{fig.15}
\end{figure}

More Shapiro steps were observed in various ranges of a broad spectrum $Fig. {\ref{fig.15}}$  of frequencies with the mesa-type of JJ like the ones shown here.

\begin{figure}[b]

\includegraphics[width=6 cm, height=4.5cm,keepaspectratio]{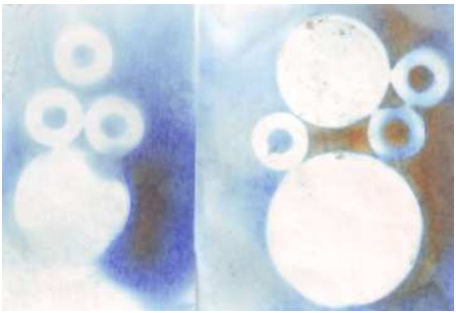}
\caption{A  JJ mesa-type of source of $THz$ waves produced the images presented below of metallic coins and washers obtained at room temperature \textbf{T = 300\:K}. The presence of tiny blue lines around the coins demonstrates presence of few $THz$ waves in the spectrum of the JJ mesa-type source.}
\label{fig.16}
\end{figure}

Several other types of Shapiro steps were observed in a broad range of frequencies with the mesa-type of JJ like the ones shown above.
Next we reproduce an image $Fig. {\ref{fig.16}}$  obtained with such a mesa-type source of $THz$ waves of coins and metallic washers.

\section{Multiple successive resistance transitions}
\begin{figure}[]
\includegraphics[width=\columnwidth]{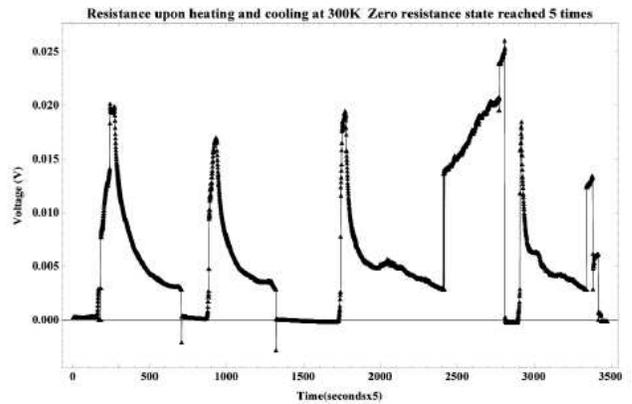}
\caption{Successive heating well above $T_{c}$ to about$\textbf{420\:K}$ of the sample and self cooling it demonstrates abrupt jumps to zero resistance state.}
\label{fig.17}
\end{figure}

Once the cooling is not necessary to observe RTS, it is enough to heat the sample and leave it to cool by itself. In an open environment this process is highly individual. With successive heating and cooling measurements of the same circuit at room temperature one can find that the sample invariably makes a very sharp jump to the smallest measured voltage of the setup. Using a Keithley Nano-Voltmeter we observed jumps to the few nano-Volts  characteristic for the zero resistance state  for the not small current of $0.05\: A$. Several such experiments were prepared with different samples which were heated up to temperatures larger than $400\:K$. In $Fig. {\ref{fig.17}}$ we plot the data  from representative measurements, which are readily verified to be different from each other as natural processes of cooling from different initial temperatures.

In a $300\:K$ environment we clearly observe successive superconductor to normal metal transitions and back to superconductor. Under repeated heating above the critical temperature and self-cooling conditions the jumps to the zero resistance state are abrupt. The slow self-cooling effect is clearly seen in longer relaxation time.The lowest measured  voltage is few nano-Volts. Here are presented some $5$ transitions to or from zero resistance state.

\section{Successive diamagnetic susceptibility transitions: $T_{c}$ in the range of 400K and above}

\begin{figure}[]
\includegraphics[width=\columnwidth]{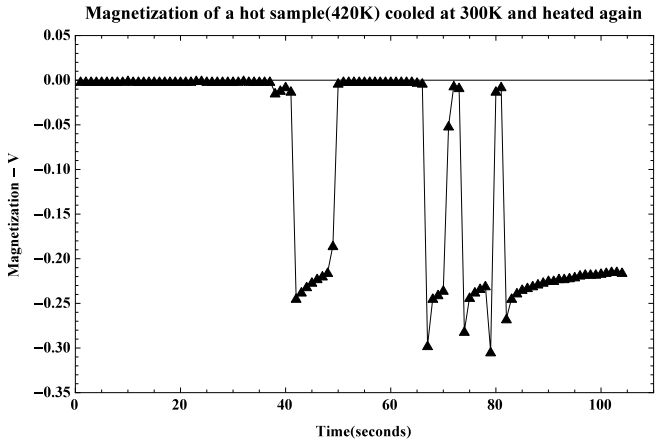}
\caption{A sample heated well above $T_{c}$ up to about$\textbf{420\:K}$  is left to cool in $\textbf{300\:K}$ environment. Whenever it reaches the critical temperature of few  tens of degrees above $\textbf{373\:K}$ it spontaneously jumps into zero resistance state. The sample is then heated again.}

\label{fig.18}
\end{figure}

Similarly, one can reproduce successive transitions to superconductor state via magnetization measurements. The only difference is that the magnitude of the signal is zero in the non-superconductor state and is maximally negative in the zero resistance state. Indeed it is visible in the next graph $Fig. {\ref{fig.18}}$ .

Exposition of samples and the effect of the magnet on diamagnetic objects is visible. Like copper,  the superconductors are attracted to the magnets. When on the surface of the magnet strong super-currents are induced on the surface of the objects creating a repulsive force acting as well as the weight.
\begin{figure}[t]
\includegraphics[width=\columnwidth,height=3cm]{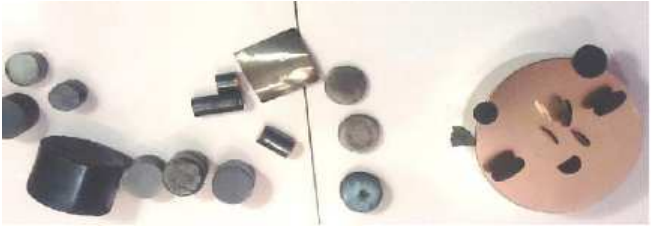}

\caption{Top view of some superconductor objects indicating different reflectivity and influence of the magnetic field making them stay upright or sideways, or stick out of the edge.}
\label{fig.19}
\end{figure}

\begin{figure}[]
\includegraphics[width=\columnwidth]{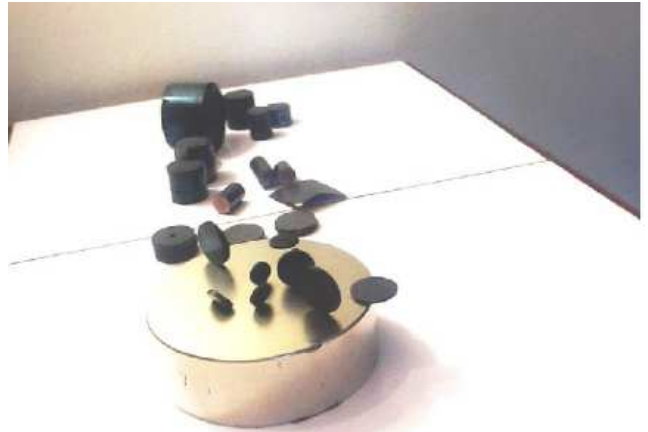}

\caption{Side view of some superconductor objects on the same magnet running away from the center or staying upright and at various angles.}

\label{fig.20}

\end{figure}

The minimal energy position depends on the shape and the size of any diamagnetic object. Left horizontally  the disks tend to spring up and stay upright as seen on the photo $Fig. {\ref{fig.19}}$. If heated above the critical temperature they remain immobile on the surface of the magnet and when cooled down to room temperature they spring up indicating a precise value of the critical temperature. This is a mechanical actuator type of sensor, which may be applied where necessary. Usually the RTS tend to reach the highest field gradient, which is at the edge of the cylinder. For this reason objects are hanging on the edge as it is seen on the following photographs $Fig. {\ref{fig.20}}$
\section{Simple test experiments}
\begin{figure}[]
\includegraphics[width=\columnwidth, height=2cm,keepaspectratio]{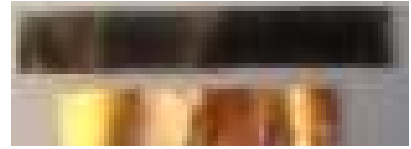}
\caption{Results of an experiment of passing $125\:A$ current via a superconductor tape (black) and a copper tape of the same size and thickness at room temperature of $300\:K$. The copper tape started burning after 5 seconds and the current was cut off. The overheated spot is seen on the left side. The tape was obtained by multi-layer deposition of about 10 microns superconductor composition on a copper tape.}
\label{fig.21}
\end{figure}

In the next experiment we connected two tapes in series: One made of copper and the other being a copper tape covered with a thin superconductor film (about $10\:microns$ thick). Passing a current of about $125\:A$ made the copper tape burn in about $5$ seconds and left the superconductor intact. A photograph of both is shown here $Fig. {\ref{fig.21}}$. Both were the same length and same width and thickness. The superconductor is longer, but the current was passed via equal pieces of both connected in series, first the superconductor and second the copper tape from the positive terminal.

\begin{figure}[]
\includegraphics[width=\columnwidth,height=2cm, keepaspectratio]{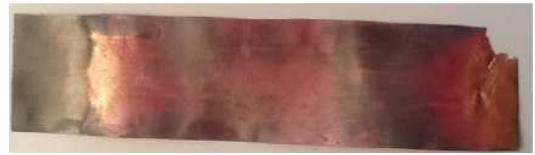}
\caption{Results of an experiment of passing $3000\:A$ current via a superconductor tape (left side of the photo) at room temperature of $300\:K$. The current evidently was passing on the superconductor surface and not along the substrate tape, the positive contact being connected to the bright left part of the tape. The flash burning was with a sharp sound and took about a fraction of a second.}
\label{fig.22}
\end{figure}

In another experiment $3000\:A$ current was passed via a different tape and reaching the critical value it "licked" the tiny superconductor layer in a flash with a short sound as seen on the photo $Fig. {\ref{fig.22}}$. The tape was obtained by multi-layer deposition of about 10 $\mu$m superconductor compound on copper tape resulting in about 80 $\mu$m total thickness of the tape.
 The tape was obtained by multi-layer deposition of about 10 $\mu$m superconductor compound on copper tape resulting in about 80 $\mu$m total thickness of the tape.

\begin{figure}[]
\includegraphics[width=\columnwidth, height=4cm, keepaspectratio]{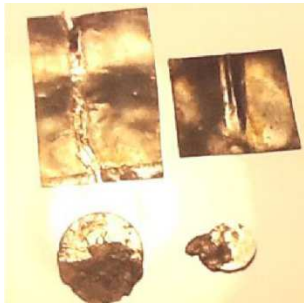}
\caption{Results of an experiment of passing a current of $2700\:A$ via the large pellet of sample \textbf{X}, $2200\:A$ via the small pellet (sample \textbf{X}), $400\:A$ via the intact tape,( it was bent upward by the Hall-Lorentz\textbf{jxB} force) and $1200\:A$ via the tape cut off by the overcritical current ( $620\:A$ were checked to be under critical) all at room temperature of $300\:K$.}
\label{fig.23}
\end{figure}
Similar experiments show that the voltage $(I -I_{c})^{n} $  dependence has a high value of n  when the tape breaks between the contacts in a tiny cut with a weak short sound(see $Fig.{\ref{fig.23}}$). Only high value of n  can explain such fast almost explosive type of heat released from the current as compared to common wires experience.

Next we give photographs of various THz sources emission in various conditions.

\begin{figure}[]
\includegraphics[width=\columnwidth]{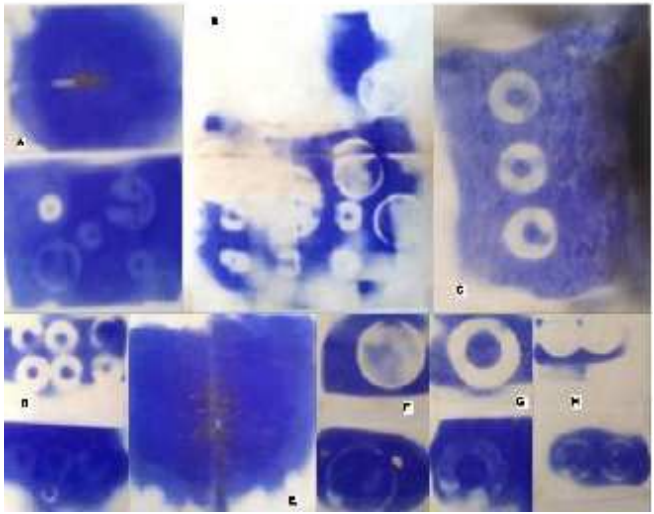}
\caption{Results of an experiment of illumination with various sources of THz waves in open air conditions (760 mm Hg pressure)  all at room temperature of $300\:K$. The real size of the images is found from the quarters( 7/8 inch) and the washers (the small are 3/8 inch and the large is 3/4 inch). The image C has a dark right side caused by discharge due to the higher then 30kV/cm field strength. The images A,D,F,G and H are composed of two different images from the two sides of one source in one experiment.  In D,F, G and H the top side is next to the tick metallic mirror, while the bottom part is emission in free air. In the top part of the image A and near the center of E one sees the traces of bunches of vortices causing brownish traces.}
\label{fig.24}
\end{figure}
All experiments of this type were performed at room temperature near 300K and at normal atmospheric pressure of 760mm Hg in open air. Some of the photographs were near a tick metallic reflecting mirror. On its surface the waves had a zero amplitude and this caused the bright images of various coins and washers. Similar images in the open space are rather dark like the image of a large flat source in photograph E and also the top of photograph A. The bottom of A include washers and coins without a mirror. The photograph B is a large source emission in the close vicinity of the tick metallic mirror causing the bright images of the absorbing metallic objects (the small washers are 3/8 inch size and the large washer is 3/4 inch). The top part of the photographs D,F,G and H are from one side of the emitter, but near the mirror, while the bottom parts of the same photographs are on the other side of the same source in open space. If a reflecting cover is placed some distance away from the source the image gets again more clear, but we do not show such one here.The image C has dark right side caused by the discharge created by the higher then 30kV/cm field strength, which was accompanied with a short sound and lightings. Values of fields strength of 40kV/cm, 80kV/cm and higher were used as pumping field in vacuum ($\sim 1 $ nanobar) for Josephson Plasma Waves (JPW) for creating THz JPWs in cuprate superconductors\cite{18}.
Various sources with power in the range below 100 W and in the range of 400W were produced and studied. A reasonable power measurement of a source about 40W was carried out both ways, by electrical and by
thermal measurements. The data for the absorption of a bag of water inside the the rectangular cavity of about 50x 40x30mm covered with a tick copper lid (its absorbed energy was about 483J as compared with the water absorbing about 4340J) show an average power generation of about 40W for 120 seconds. The result is a THz "flash bulb" of 40W emitting broad band THz waves very effectively.
\section{ Acknowledgements }
The author would like to thank all his colleagues from the University of Sofia and the other academic institutions mentioned in the cited papers below, with whom he started working on superconductivity in the Spring of 1987. He also thanks all his colleagues and friends and is indebted in particular to his Ohio State University colleagues.

\end{document}